\documentclass[twocolumn]{aastex63}
\usepackage{amsmath}

\accepted{November 9, 2020}
\received{May 9, 2020}

\shorttitle{SCORCH.III. Analytical Models and Clumping Factors}
\shortauthors{Chen et al.}

\begin{document}

\title{SCORCH. III. Analytical Models of Reionization with Varying Clumping Factors}

\correspondingauthor{Nianyi Chen}
\email{nianyic@andrew.cmu.edu}

\author[0000-0001-6627-2533]{Nianyi Chen}
\affiliation{McWilliams Center for Cosmology, Department of Physics, Carnegie Mellon University, Pittsburgh, PA 15213}

\author[0000-0002-4586-7586]{Aristide Doussot}
\affiliation{Sorbonne Universit{\'e}, Observatoire de Paris, PSL Research University, CNRS, LERMA, F-75014 Paris, France}

\author[0000-0001-6778-3861]{Hy Trac}
\affiliation{McWilliams Center for Cosmology, Department of Physics, Carnegie Mellon University, Pittsburgh, PA 15213}

\author[0000-0001-8531-9536]{Renyue Cen}
\affiliation{Department of Astrophysical Sciences, Princeton University, Princeton, NJ 08544}


\begin{abstract}
In the Simulations and Constructions of the Reionization of Cosmic Hydrogen (SCORCH) project, we compare analytical models of the hydrogen ionization fraction with radiation-hydrodynamic simulations. We derive analytical models of the mass-weighted hydrogen ionization fraction from the local ionization balance equations as a more accurate alternative to the widely adopted model based on the volume filling factor. In particular, our model has a recombination term quadratic in the ionization fraction, which is consistent with the two-body interaction nature of recombination. Then, we use the radiation-hydrodynamic simulations to study the clumping factors needed to solve the analytical equations, and provide accurate fitting functions. We find that the ionized hydrogen clumping factors from our radiative transfer simulations are significantly different than those from other simulations that use a uniform photoionization background. In addition to redshift dependence, we also see the dependence of ionized hydrogen clumping factor on ionization fraction, and we incorporate this into our fits. We calculate the reionization histories using our analytical models and clumping factors and compare with widely adopted models, and all of our models achieve $<7\%$ difference from simulation results while the other models have $>20\%$ deviations. The Thomson optical depths from reionization calculated from our analytical models result in $<5\%$ deviation from simulations, while the previous analytical models have $>20\%$ difference in and could result in biased conclusions of the IGM reionization.

\end{abstract}


\section{Introduction}
\label{sec:introduction}

The Epoch of Reionization (EoR) is a period when the first stars, galaxies and quasars emit UV photons and ionize the neutral hydrogen in the universe. These photons have large impact on the state and temperature of the baryonic gas through photoionization and photoheating and hence also affect the structure formation in the late universe. Therefore, we can gain information about the first generation of luminous sources and the status of the IGM, as well as constrain astrophysics and cosmology by tracing the detailed history of reionization. 

Due to the limited observational data at this relatively high redshift, knowledge about the sources of ionizing photons and the evolution of the IGM is still incomplete, but some progress has already been made. For example, high-redshift galaxy observations show that galaxies most likely provided the bulk of the ionizing photons \citep[e.g][]{Bouwens2015b, Finkelstein2016a, Livermore2017a}, but quasars could make some contribution towards the end of reionization \citep[e.g][]{Madau2015a}. \cite{Planck2018a} recently inferred a Thomson optical depth $\tau= 0.054 \pm 0.007$ from measurements of the CMB temperature and polarization angular power spectra, implying a late reionization midpoint at redshift $z = 7.7 \pm 0.6$ \citep[e.g.][]{Glazer2018a}. \cite{Becker2015a} find evidence of a dark Ly$\alpha$ trough extending down to z $\sim$ 5.5 in the spectrum of a high-redshift quasar, suggesting that reionization could have ended at $z < 6$ \citep[e.g.][]{Keating2019a,Nasir2020a}, later than previously assumed.

On the theoretical side, there are three main approaches to study EoR. The most accurate and expensive are the cosmological simulations combining N-body, hydro, and radiative transfer (RT) algorithms to solve the coupled evolution of the dark matter, baryons,
and radiation \citep[e.g.][]{Trac2008a, Gnedin2014a, Norman2015a, Semelin2017a, Finlator2018a, Doussot2019a}. On the next level of accuracy, there are semi-analytical/numerical methods providing an approximate and efficient approach to solve both the spatial and temporal evolution of the reionization process. They are especially useful for making mock observations on large scales \citep[e.g.][]{Furlanetto2004a, Zahn2007a,Alvarez2009a, Santos2010a, Mesinger2011a, Battaglia2013a}. The most convenient but least accurate are fast analytical calculations and models, which are preferred for exploring the large parameter space in forecasting or inference studies \citep[e.g.][]{Madau1999a, Miralda-Escude2000a, Barkana2004a, Kaurov2014a}. 

One of the most commonly used analytical model is a differential equation for the time evolution of the volume filling factor of ionized hydrogen (HII) by \cite{Madau1999a} (hereafter M99). This model has recently been used to constrain the reionization history and infer properties of the radiation sources such as the ionizing emissivity and radiation escape fractions \citep[e.g.][]{McQuinn2011a,Haardt2012a,Kuhlen2012a,Bouwens2015a, Robertson2015a, Price2016a,  Madau2017a, Ishigaki2018a}. Despite its wide-spread use, there are several concerns to this model: first, it was deriving Str{\"o}mgren sphere analysis under the simplifying assumption of isolated HII regions; furthermore, it has generally been applied assuming a constant clumping factor in the recombination term, which is not true in reality.

To study and use the analytical models of reionization fraction, one would often need to make use of the clumping factors in order to simplify the calculations. Among different clumping factor definitions, the ionized hydrogen clumping factor is of most interest in the literature. The ionized hydrogen clumping factor accounts for the distribution of ionized hydrogen in the IGM. Most numerical and semi-analytical simulations of EoR choose to incorporate constraints on the clumping factors into their simulations \citep[e.g.][]{Bouwens2015a,Robertson2015a,Ishigaki2015a,Greig2017a}. In addition, many recent studies tend to directly consider this quantity as independent of the redshift and roughly constant, which is yet unproven. On the contrary, some studies have discussed its likely variability \citep{Gorce2018a} while others tried to look for systematic errors in the computation or emphasize a possible scale-dependency \citep{Kaurov2015a}. 
 
The Simulations and Constructions of the Reionization of Cosmic Hydrogen (SCORCH) project is designed to provide radiation-hydrodynamic simulations, theoretical predictions, and mock observations to facilitate more accurate comparisons with current and future observations. In SCORCH I \citep{Trac2015a}, we probe the connection between observed high-redshift galaxies and simulated dark matter halos to better understand the primary source of ionizing photons. By abundance matching galaxy UV luminosities to halo mass accretion rates, we construct a fiducial model for the galaxy luminosity functions that can be extrapolated to fainter magnitudes and higher redshifts. Building on this work,  \cite{Price2016a} use both parametric and non-parametric statistical methods to constrain the radiation escape fraction $f_{\rm{esc}}(z)$ from high-redshift galaxies using HST and Planck observations. Their inferred results favor increasing $f_{\rm{esc}}$ towards higher redshift in an approximately power-law relation. With a better understanding of the evolving abundance of high-redshift galaxies and the production of ionizing radiation, in SCORCH II \citep{Doussot2019a} we run and analyze three radiation- hydrodynamic simulations with the same fiducial galaxy luminosity functions, but different radiation escape fraction models. The simulations are designed to have the same $\tau_e = 0.06$ and similar midpoints of reionization $7.5 < z < 8$, but with different ionization histories. Recently, \cite{DAloisio2019a} have also used these simulations to study the heating of the intergalactic medium (IGM) by hydrogen reionization.

In this paper, we derive an analytical model for global hydrogen ionization fraction from the local ionization balance equation, and study different clumping factors related to reionization with the RadHydro simulations in the context of these analytical models. We highlight the dependency of the clumping factors not only on the redshift, but also on the ionization fraction. We also compare the current analytical models of reionization and our own models derived from simulation data in terms of the solved reionization history and the resulting Thomson optical depth. 
This paper is organized as follows: in Section \ref{sec:Method}, we first summarize the RadHydro simulations used to conduct our analysis. Then we review the widely adopted M99 model for reionization history before proposing and deriving our own analytical model, and we end this section by describing details about our measurements of the clumping factors from the simulation. In Section \ref{sec:results}, we first present simulation data and fits for ionized hydrogen clumping factors, recombination clumping factors, ionization clumping factors and total hydrogen clumping factors. Then we compare the reionization histories solved from different analytical models and clumping factors. Finally, we show the comparison between the Thomson optical depths calculated from different analytical models and the simulation results. We adopt the following cosmological parameters in the simulations: $\Omega_\text{m} = 0.30$, $\Omega_\Lambda = 0.70$, $\Omega_\text{b} = 0.045$, $h = 0.70$, $n_\text{s} = 0.96$ and $\sigma_8 = 0.80$.

\section{Method}
\label{sec:Method}

\subsection{Radiation-hydrodynamic Simulations}
\label{sec:scorch}

\begin{deluxetable*}{lCCCCCCCCC}
\label{tab:scorch}
\tablewidth{\textwidth}
\tablecaption{RadHydro simulations}
\tablehead{
\colhead{Model} & \colhead{$L\ [h^{-1}\text{Mpc}]$} & \colhead{$N_\text{dm}$} & \colhead{$N_\text{gas}$} & \colhead{$N_\text{RT}$} &
\colhead{$f_8$} & \colhead{$a_8$} & \colhead{$\tau$} & \colhead{$z_\text{mid}$} &  \colhead{$\Delta_\text{z}$}
}
\startdata
Sim 0 & 50 & $2048^3$ & $2048^3$ & $512^3$ & 0.15 & 0 & 0.060 & 7.95 & 4.68 \\
Sim 1 & & & & & 0.13 & 1 & 0.060 & 7.91 & 5.45  \\
Sim 2 & & & & & 0.11 & 2 & 0.060 & 7.83 & 6.54
\enddata
\end{deluxetable*}

In SCORCH II \citep{Doussot2019a}, we present three radiation-hydrodynamic simulations with the same cosmic initial conditions, same galaxy luminosity functions, but with different radiation escape fraction models. The simulations are run with the RadHydro code which combines N-body and hydrodynamic algorithms \citep{Trac2004a} with an adaptive raytracing algorithm \citep{Trac2007a}. It directly and simultaneously solves collisionless dark matter dynamics, collisional gas dynamics, and radiative transfer of ionizing photons. The ray tracing algorithm uses adaptive splitting and merging to improve resolution and scaling. This code has been previously used to simulate both hydrogen and helium reionization \citep[e.g.][]{Trac2008a,Battaglia2013a,LaPlante2016a}. 

The RadHydro simulations have $2048^3$ dark matter particles, $2048^3$ gas cells and up to 12 billion adaptive rays in a $50\ h^{-1} \text{Mpc}$ comoving box. We track five frequencies (15.7, 21.0, 29.6, 42.9, 74.1 eV) above the 13.6 eV hydrogen ionizing energy. We then compute the incident radiation flux and use it in the computation of the photoheating and photoionization rates needed by the nonequilibrium solvers to solve the ionization and energy equations. The same initial conditions, generated at a starting redshift of 300, is used in all of the simulations. All three simulations are run down to redshift $z = 5.5$.

Using an updated subgrid approach to model the radiation sources, we are both able to populate dark matter halos with galaxies, by matching the galaxy luminosity functions, and accurately compute the spatial distribution of ionizing sources. Following SCORCH I \citep{Trac2015a}, the luminosity-accretion rate $L_{\text{UV}}(\dot{M})$ relation is inferred from the halo mass accretion rate and the abundance matching performed by equating the number density of galaxies to the number density of halos.

To generate halo and galaxy catalogs, a particle- particle-mesh ($\text{P}^3\text{M}$) N-body simulation with $3072^3$ dark matter particles is run using a high- resolution version of the same initial conditions as the RadHydro simulations. A hybrid halo finder is run on the fly every 20 million cosmic years to locate dark matter halos and build merger trees. With a particle mass resolution of $3.59\times 10^5 h^{-1} M_\odot$, we can reliably measure halo quantities such as mass and accretion rate down to the atomic cooling limit.

These simulations are consistent with the latest Planck observations \citep{Planck2016a,Planck2018a}, as they have been designed to have fixed Thomson optical depth $\tau_e = 0.06$. They start with the same initial conditions and have the same galaxy populations, but use different radiation escape fraction models $f_{\text{esc}}$. Following \cite{Price2016a}, we choose a two-parameter single power-law:
\begin{equation}
    f_{\text{esc}}(z) = f_8 \left( \frac{1+z}{9}\right)^{a_8} ,
\end{equation}
where $f_8$ is the value of the escape fraction at $z = 8$
and $a_8$ is the exponent that change between our three
simulations. With our three runs we test $a_8$ = 0, 1, and
2.

Table \ref{tab:scorch} summarizes the parameters for the three RadHydro simulations. $z_{\rm mid}$ is the redshift at which $50\%$ of the hydrogen is ionized, and $\Delta_{\rm z}$ is the redshift interval between $5\%$ and $95\%$ ionization. From the table, we see that the main difference in the three simulations is the reionization histories resulted from different treatment of the escape fraction. Sim 0 has constant $f_{\rm esc}$ and reionization starts latest, but ends earliest out of the three models. Sim 1 has $f_{\rm esc}(z)$ varying linearly with $1+z$ and is an intermediate model. Sim 2 has $f_{\rm esc}(z)$ varying quadratically and reionization starts earliest, but ends latest. For more details on how different models affect other aspects of reionization, please see SCORCH I and II \citep{Trac2015a,Doussot2019a}.

\subsection{Volume Filling Factor Model}
\label{sec:madau}

We begin the discussion of analytical models of reionization history by reviewing the reference model proposed by \cite{Madau1999a} (we will refer to this model as M99 throughout the paper), which has been widely used in analytical calculations \citep[e.g.][]{Bolton2007a,Kuhlen2012a,Bouwens2015a,Price2016a}:
\begin{equation}
\label{eq:m99}
    \frac{dQ_{\text{HII}}}{dt} = \frac{\langle \dot{n}_\gamma \rangle_{\rm{V}}}{\langle n_{\rm{H}} \rangle_{\rm{V}}} - \frac{Q_{\text{HII}}}{ \bar{t}_{\rm{rec}}} ,
\end{equation}
where $Q_{\rm{HII}}$ is the volume filling factor of ionized hydrogen, $\langle n_{\text{H}} \rangle_{\rm{V}}$ is the volume-averaged total hydrogen number density, $\langle \dot{n}_\gamma \rangle_{\rm{V}}$ is the volume-averaged photon production rate, and the effective recombination time is given by 
\begin{equation}
    \bar{t}_{{\rm{rec}}} = \frac{1}{(1+Y/4X)\alpha (T_0) \langle n_{\text{H}}\rangle_{\rm{V}} C} ,
 \end{equation}
 where $X=0.76$ and $Y=0.24$ are the hydrogen and helium mass fractions, respectively, $C$ is the clumping factor, $\alpha$ is the recombination coefficient and $T_0$ is the temperature of the IGM at mean density, fixed to be $T_0 = 2\times10^4K$ throughout this paper in order to match the expectations from star-forming galaxy spectra \citep[e.g.][]{Hui2003a,Trac2008a}. 

It is worth noting that 
Equation (2)
is derived from a constant density hypothesis. It thus implies that the mass-weighted and volume-weighted ionization fractions are equal.

\subsection{Mass-weighted Ionization Fraction Model}
\label{sec:balance}

As an alternative to the volume filling factor, in this paper we propose an analytical model for solving the mass-weighted ionization fraction, because it comes from the widely used reionization balance equations and is frequently used when calculating observables. The mass-weighted global ionization fraction can be calculated as the ratio between volume-weighted number densities:
\begin{align}
\label{eq:mass-weighted}
    \langle x_\mathrm{HII}\rangle_\mathrm{M} & = \frac{\sum \rho_{\mathrm{H},i}x_{\mathrm{HII},i}}{\sum \rho_{\mathrm{H},i}}
    = \frac{\sum \rho_{\mathrm{HII},i}}{\sum \rho_{\mathrm{H},i}}
     \\ \nonumber 
    & = \frac{\langle \rho_\mathrm{HII} \rangle_\mathrm{V}}{\langle \rho_\mathrm{H} \rangle_\mathrm{V}} = \frac{\langle n_\mathrm{HII} \rangle_\mathrm{V}}{\langle n_\mathrm{H} \rangle_\mathrm{V}},
\end{align}
where the summation is over all the cells in the simulation, and we have used the relation $\rho_{\rm H} = m_{\rm H}n_{\rm H}$ in each of the Eulerian cells.

In order to obtain a self-consistent, rigorous derivation of the global reionization equation for the mass-weighted ionization fraction, we start with the local ionization balance equation of hydrogen \citep[e.g.][]{Gnedin1997a}:
\begin{equation}
\label{eq:local_balance}
    \frac{dn_{\text{HII}}}{dt} = \Gamma n_{\text{HI}}  + \gamma_{\rm coll}n_{\rm e} n_{\rm HI}-\alpha(T)n_{\rm{e}} n_{\text{HII}} - 3Hn_{\text{HII}} ,
\end{equation}
where $n_{\text{HII}}$ is the physical number density of ionized hydrogen, $n_{\rm{e}}$ is the physical number density of free electrons, $\gamma_{\rm coll}$ is the collisional ionization rate,  $\Gamma$ is the photoionization rate, $\alpha$ is the recombination coefficient and $H$ is the Hubble parameter. The right-hand side contains four effects that govern the local ionization of hydrogen: the first two terms are the photoionization and collisional ionization respectively, which increases ionized hydrogen number density, the third term is the recombination which decreases the ionized hydrogen number density, while the last term accounts for the decrease in physical number density due to the universal expansion. Because the collisional ionization is insignificant compared with photoionization in the low-density IGM regions we are interested in, in subsequent derivations we will drop the second term and focus on photoionization only.

Taking volume-weighted average on both sides of
Equation \ref{eq:local_balance} and dividing both sides by $\langle n_{\text{H}} \rangle_{\rm V}$,
we obtain the global equation for the mass-weighted ionization fraction:
\begin{equation}
\label{eq:global_balance}
     \frac{d\langle x_{\text{HII}}\rangle_{\rm{M}}}{dt} = \frac{\langle \Gamma n_{\text{HI}} \rangle_{\rm{V}}}{\langle n_{\text{H}} \rangle_{\rm{V}}} - \frac{\langle\alpha(T)n_{\rm{e}} n_{\text{HII}}\rangle}{\langle n_{\text{H}} \rangle_{\rm{V}}},
\end{equation}
where the left-hand side follows from Equation \ref{eq:mass-weighted}.

Instead of the volume filling factor $Q$, we focus on the evolution of  $\langle x_{\text{HII}} \rangle_{\rm{M}}$, as this quantity is more relevant for calculating observables such as the Thomson optical depth.  Note that the fourth, universal expansion term on the right-hand side of Equation \ref{eq:local_balance} has been cancelled by the same term on the left-hand side 
that results from time derivative of the denominator
of Equation \ref{eq:mass-weighted}.

One issue with solving for global reionization directly with Equation \ref{eq:global_balance} is that the last term is difficult to compute. Thus, the recombination term is often parametrized by the clumping factor for quick calculations. If we relate the free electron number density $n_{\rm{e}}$ to the density of ionized hydrogen $n_{\text{HII}}$, assuming helium is singly ionized, by:
\begin{equation}
    n_{\rm{e}} = \left( 1+\frac{Y}{4X} \right)n_{\text{HII}} ,
\end{equation}
we can then rewrite Equation \ref{eq:global_balance} in terms of the ionization and recombination clumping factors as:

 \begin{align}
\label{eq:Gamma}
       \frac{d\langle x_{\text{HII}}\rangle_{\rm{M}}}{dt} 
    &= C_{\text{I}}\langle \Gamma \rangle_{\rm{V}} (1-\langle x_{\text{HII}} \rangle_{\rm{M}})\\ 
    &- C_{\text{R}}\langle \alpha(T) \rangle_{\rm{V}}\langle n_{\text{H}} \rangle_{\rm{V}} \left(1+\frac{Y}{4X} \right)\langle x_{\text{HII}}\rangle_{\rm{M}}^2,   \nonumber
\end{align}
where $C_{\text{I}}$ is the photoionization clumping factor  \citep[e.g.][]{Kohler2007a} that occurs due to spatial fluctuations in the radiation field:
\begin{equation}
\label{eq:CI}
    C_{\text{I}} = \frac{\langle \Gamma n_{\text{HI}} \rangle_{\rm{V}}}{\langle \Gamma \rangle_{\rm{V}} \langle n_{\text{HI} } \rangle_{\rm{V}}},
\end{equation}
 and  $C_{\text{R}}$ is the recombination clumping factor: 
\begin{equation}
\label{eq: Ceff}
    C_{\text{R}} = \frac{\langle \alpha(T)n_{\text{HII}} n_e \rangle_{\rm{V}}}{\langle\alpha(T)\rangle_{\rm{V}}\langle n_{\text{HII} } \rangle_{\rm{V}} \langle n_{e} \rangle_{\rm{V}}}.
\end{equation}

Here we use the most general way of defining the clumping factor, namely that we take the spatial temperature variation into account when taking the global average of the recombination rate instead of assuming a constant temperature.

So far we have defined the clumping factors and shown how the global ionization fraction can be iteratively solved from the differential equation once the clumping factors are known. Before discussing in more detail the measurements of clumping factors, we would like to point out that the ionization rate $\Gamma$ is often difficult to compute without directly using the simulations. A more convenient quantity to use in the photoionization term is the ionizing photon production rate, $\dot{n_\gamma}$. This quantity can be computed from galaxy luminosity functions and the escape fraction without running
radiative transfer simulations. If we use the photon production rate $\dot{n_\gamma}$ in the place of photoionization rate $\Gamma$, Equation \ref{eq:Gamma} becomes:
\begin{align}
\label{eq:ndot}
       \frac{d\langle x_{\text{HII}}\rangle_{\rm{M}}}{dt} 
    &= \frac{\langle\dot{n}_{\gamma}\rangle_{\rm{V}}}{\langle n_{\text{H}}\rangle_{\rm{V}}}\\ 
    &-\mathbf{C_{\text{R}}\langle \alpha(T) \rangle_{\rm{V}}}\langle n_{\text{H}} \rangle_{\rm{V}} \left(1+\frac{Y}{4X} \right)\langle x_{\text{HII}}\rangle_{\rm{M}}^2 .  \nonumber
\end{align}
Notice that only the first term on the right-hand side changes, and there is no need of an ionization clumping factor $C_{\rm I}$ in this case. In the following sections, we will use both Equation \ref{eq:Gamma} and \ref{eq:ndot} to solve for the reionization history and will evaluate the performance of both.

\subsection{Clumping Factors}
\label{sec:clumping}

We have defined clumping factors and shown their usefulness in solving for reionization history in the previous section. Now we want to provide details about the physical meaning of clumping factors and how we calculate them from our simulations.

In simulation subgrid modeling and global analytical models, clumping factors are used to account for the excess of recombination or photoionization,
due to fluctuations in gas density and radiation field, respectively,
when solving for ionization fractions in the IGM. When calculating clumping factors from the simulations, the intra-halo gas is often excluded, because the ionization within the halo is already accounted for in the escape fraction. 
Since we model $f_{\rm{esc}}$ explicitly in our RadHydro simulations, we also exclude the halo gas in our clumping factor computation, and we do so by applying empirical upper limits on the gas overdensity. If the overdensity of a region is greater than this limit, the region is considered to be a part of the intra-halo gas and is not taken into account during the computation.

In order to find a robust functional form and to make our models broadly applicable, we choose to study clumping factors under three different density upper limits. We use $\Delta=50$, $\Delta=100$ and $\Delta=200$ in units of the global mean gas density as our three density cuts, and the resulting clumping factors are named $C_{\Delta<50}$, $C_{\Delta<100}$ and $C_{\Delta<200}$, respectively. The clumping factors are computed while the RadHydro simulations are running instead of in post-processing. Doing the calculations at many time steps instead of for a few saved snapshots allows us to better study the redshift evolution.

We have not done extensive convergence tests for the clumping factors due to limited computational resources. In addition to the fiducial high-resolution simulations, we have run low-resolution and medium-resolution versions that have 4 and 2 times lower spatial resolution, respectively. However, we have not attempted ultra-high resolution simulations with smaller volumes as our current box is already about the minimum size required to accurately capture larger ionized regions. Furthermore, the calculations and models for physical processes such as cooling, star formation, and feedback are quite sensitive to resolution especially in high-density regions. While we do see smaller variations in the clumping factors between the two highest-resolution runs compared to the two lowest-resolution runs, it is difficult to do a fair comparison for the reasons stated earlier. In the following sections, we provide estimated uncertainties due to resolution effect from the differences between our mid-resolution and high-resolution runs. The readers can treat our clumping factor measurements as lower limits, and use the upper limits provided in the figures as a guidance for higher-resolution results.


\section{Results}
\label{sec:results}

Having formulated how to use the clumping factors to solve for reionization history and shown how we define and calculate the clumping factors in the IGM from the RadHydro simulations, in this section we will provide measurements, fitting formulas and parameters for the clumping factors. We will also show that with our fits and our models for the global reionization balance equation, we can recover the evolution of ionized hydrogen accurately in comparison to direct radiation hydrodynamics simulations.

\subsection{Total Hydrogen Clumping Factor}
\label{sec:Total}

\begin{figure}[t]
\includegraphics[width=\hsize]{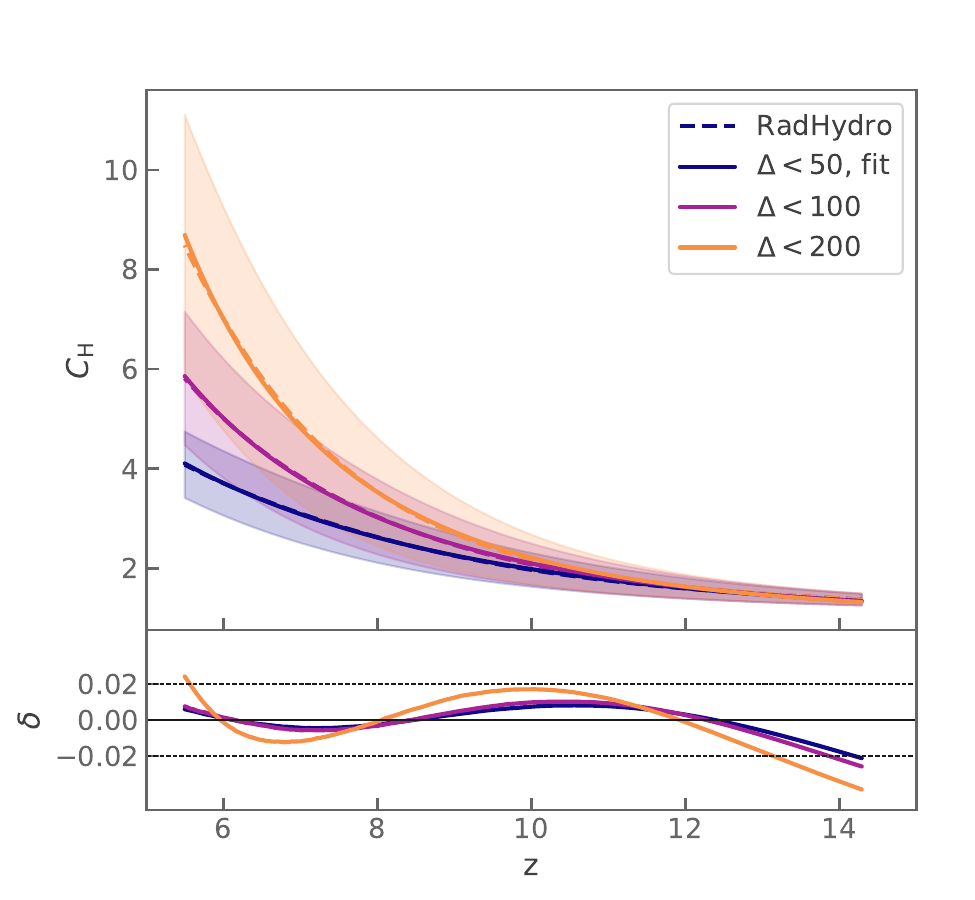} 
\caption{The total hydrogen clumping factor $C_{\text{H}}$ as a function of the redshift for the three definitions given by the simulation (\textbf{dashed}) and the fitting (\textbf{solid}). The shaded bands show the uncertainty estimation from resolution effect. The relative errors are shown in the lower panels. Fitting function is given in Equation \ref{eq:cH_fit} and parameters are given in Table \ref{tab:CH}.}
\label{fig:cH} 
\end{figure}

\begin{deluxetable}{lCCCCC}
\label{tab:CH}
\tablewidth{\textwidth}
\tablecaption{$C_{\rm H}$: Total Hydrogen Clumping Factor}
\tablehead{
 & & $C_{\rm H}$\\
\hline
\colhead{Model} & \colhead{$C_0$} & \colhead{$\alpha_0$} &  \colhead{$\alpha_1$}   &\colhead{$\delta_{\text{rms}}$} & \colhead{$\delta_{\text{max}}$}
}
\startdata
$C_{\text{H},\Delta<50}$ &1.62 & -2.27 & -0.107  & 0.4\%& 2.1\%\\
$C_{\text{H},\Delta<100}$ &2.02 & -2.89 & -0.079 & 0.5\% & 2.6\%\\
$C_{\text{H},\Delta<200}$ &2.53 & -3.55 & -0.056  & 1.1\% & 3.9\%\\
\enddata
\end{deluxetable}

\begin{deluxetable}{lCCC}
\label{tab:CH_range}
\tablewidth{\textwidth}
\tablecaption{Uncertainty Range Fitting for $C_\text{H}$}
\tablehead{
 \colhead{Model} & \colhead{$C_0$} & \colhead{$\alpha_0$} & \colhead{$\alpha_1$}
}
\startdata
$C_{\text{H},\Delta<50}$ & [1.1,2.1] & [-2.6,-2.1] & [-0.064,-0.13]\\
$C_{\text{H},\Delta<100}$ & [1.3,2.8] & [-3.2,-2.8] & [-0.017,-0.11] \\
$C_{\text{H},\Delta<200}$ & [1.5,3.6] & [-3.7,-3.5]  & [0.030,-0.091]
\enddata
\end{deluxetable}

The total hydrogen clumping factor $C_\text{H}$ is calculated from all hydrogen gas, both neutral and ionized, from the simulation. It is defined as:
\begin{equation}
    C_{\text{H}} = \frac{\langle n_\text{H}^2 \rangle_{\rm{V}}}{\langle n_\text{H} \rangle^2_{\rm{V}}}.
\end{equation}
The total hydrogen clumping factor is not used in the ionization equations mentioned in the previous section, but we provide our measurements and fits for it here for completeness.

Figure \ref{fig:cH} shows the evolution of the total hydrogen clumping factor $C_\text{H}$ for our three definitions of $C_\text{H}$ given by the simulation. The total hydrogen clumping factor increases with time, in agreement with the commonly acknowledged fact that the collapse gas fraction increases with time during the EoR. Furthermore, when we apply a higher upper limit in density, we are including more gas around the halos into our calculation. As the gas density around halos grows at lower redshift, we would expect a higher $C_{\text{H}}$ for the high-density-cut models such as $\Delta<200$, as can be seen in the plot.
To show the uncertainties in $C_{\text{H}}$ from numerical resolution, we also plot in Figure \ref{fig:cH} the estimated error bands of the clumping factors based on the differences between the mid-resolution and high-resolution simulations. There is a $\sim 20\%$ difference between different resolutions.

As both ionized and neutral hydrogen atoms are considered indifferently, $C_\text{H}$ does not depend on the ionization fraction $x_{\text{HII}}$, so we consider its evolution only as a function of the redshift. Moreover, because of its independence of $x_{\text{HII}}$, the value of the total hydrogen clumping factor is almost the same for Sims 0, 1, and 2. It does, however, depend on the maximum overdensity $\Delta$ where measurements are taken. Hence, we fit three sets of parameters for different maximum overdensities but not different simulations.

For each density cut, we fit the hydrogen clumping factor by a single power-law with running exponent. The fitting formula is:
\begin{equation}
\label{eq:cH_fit}
    C_{\text{H}}(z) = 1+C_0 \left( \frac{1+z}{9} \right)^{\alpha(z)}
\end{equation}
where $C_0$ is a constant controlling the overall amplitude, $\alpha$ is a linear function in redshift with $\alpha (z) = \alpha_0 + \alpha_1 (z-8)$.

To fit this function to our simulation data, we use the basin-hopping algorithm \citep{Wales1997a} and minimize the relative error:
\begin{equation}
\label{eq:error}
    \epsilon = \sum_{z = z_0}^{z_{\rm{n}}} \frac{|C_{\text{sim}} - C_{\text{fit}}|}{C_{\text{sim}}}.
\end{equation}
Here we choose $z_0=14$ and $z_{\rm{n}}=5.5$ to be the redshift range of our fitting for $C_{\rm H}$. We optimize three sets of parameters for the three density threshold measurements. 

The results of our fitted parameters are shown in Table \ref{tab:CH}. For low redshifts ($z<10$), the $\alpha_0$ term dominates in the exponent as $\alpha_1$ in the linear term is relatively small, and high-$\Delta$ models have more negative $\alpha_0$ which leads to a steeper slope, which is physically due to the coincidence of the rapid rise of mass fraction in halos that are both sources and sinks of reionization process at the last stage of the reionization. As we go to higher redshifts, the linear term in the exponent begins to dominate, and the values for three density cuts cross with each other. We also fit the upper- and lower-limits of the clumping factors estimated from different resolutions and provide fitting parameters in Table \ref{tab:CH_range}, in order to show how the range of parameters corresponds to the range in clumping factors. The fitted curves are plotted alongside the simulation data in Figure \ref{fig:cH}, and their differences are shown in the lower panel.

\subsection{Ionized Hydrogen Clumping Factor}
\label{sec:sink}

\begin{figure*}[t]
\includegraphics[width=1.0\hsize]{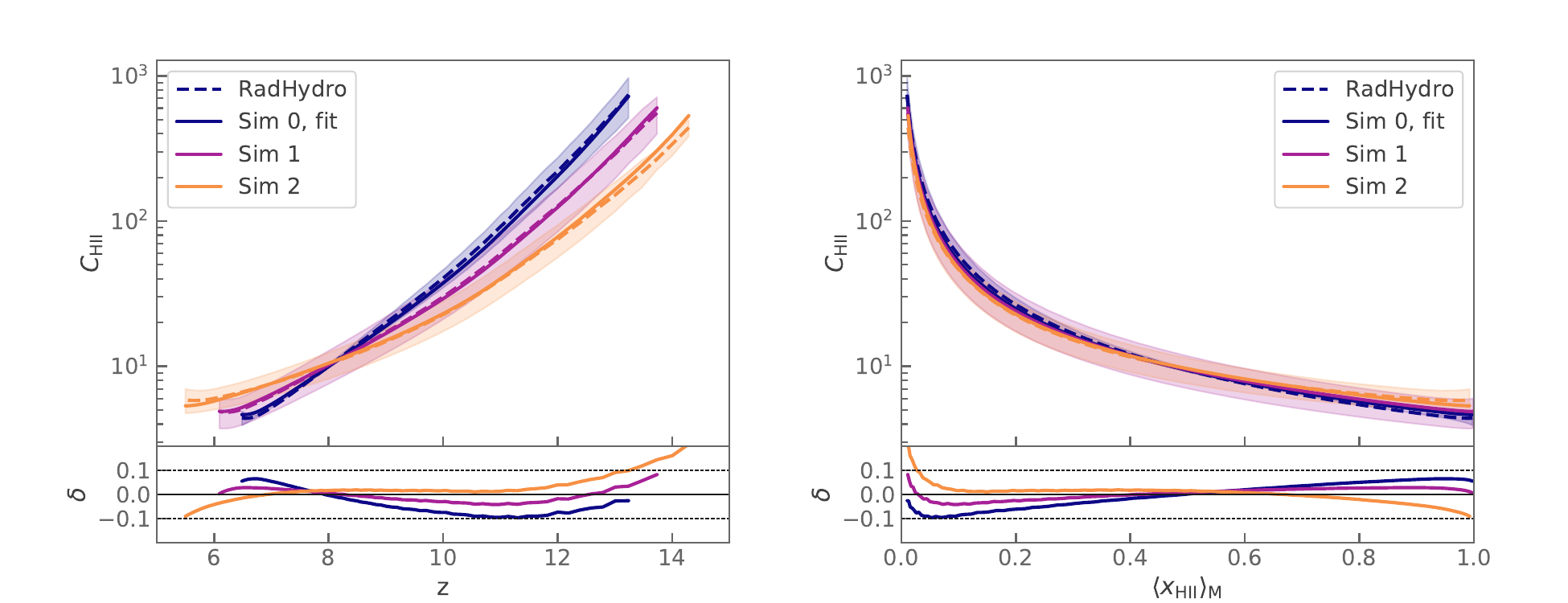}
\caption{The ionized hydrogen clumping factors from the three RadHydro simulations and the fitted curves for the $\Delta<100$ density cut. We show the simulation data (\textbf{dashed}) and estimated resolution uncertainties (shaded regions) along with the fitted data from Equation \ref{eq:cHII_fit} (\textbf{solid}) in for Sims 0 (\textbf{blue}), 1 (\textbf{purple}) and 2 (\textbf{orange}). The lower panels shows the deviations of the fits from the simulation data.} 
\label{fig:cHII}
\end{figure*}

\begin{figure*}[t]
\includegraphics[width=1.0\hsize]{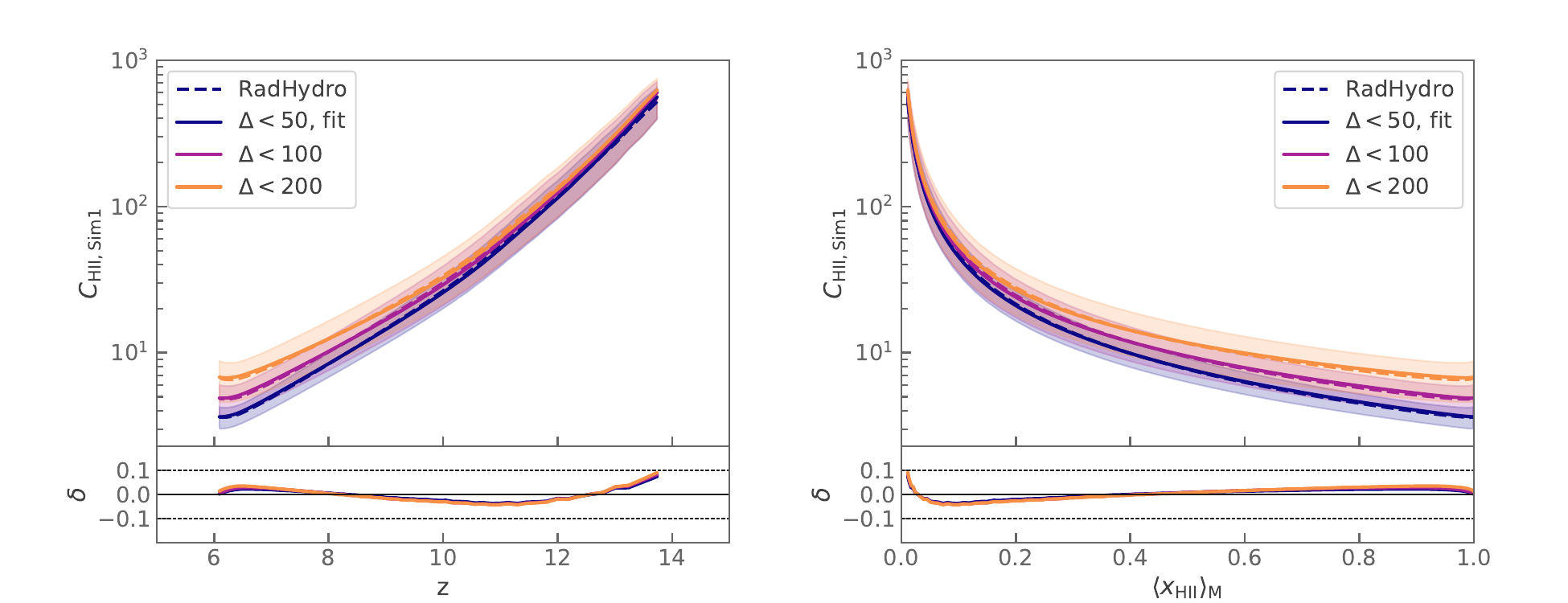}
\caption{The ionized hydrogen clumping factors with estimated uncertianties from Sim 1 for different density cuts $\Delta<50$ (\textbf{blue}), $\Delta<100$ (\textbf{purple}) and $\Delta<200$ (\textbf{orange}). With a higher density cut, we are including more high-density gas around the halos and the overall clumpiness is higher.} 
\label{fig:cHII_rho}
\end{figure*}

\begin{deluxetable}{lCCCCCC}
\label{tab:CHII}
\tablewidth{\textwidth}
\tablecaption{$C_\text{HII}$: Ionized Hydrogen Clumping Factor}
\tablehead{
 & & & $C_\text{HII}$\\
\hline
\colhead{Model} & \colhead{$C_0$} & \colhead{$\alpha_0$} & \colhead{$\alpha_1$} &\colhead{$\beta_0$} & \colhead{$\delta_{\text{rms}}$} & \colhead{$\delta_{\text{max}}$} 
}
\startdata
$C_{\text{HII},\Delta<50}$ & 2.07 & -1.43 & -0.213 & -1.32 & 5.0\% & 9.7\%\\
$C_{\text{HII},\Delta<100}$ & 2.69 & -1.92 & -0.196 & -1.33  & 2.2\% & 8.3\%\\
$C_{\text{HII},\Delta<200}$ & 3.48 & -2.50  & -0.185 & -1.34 & 3.1\% & 21\%
\enddata
\end{deluxetable}

\begin{deluxetable}{lCCCC}
\label{tab:CHII_range}
\tablewidth{\textwidth}
\tablecaption{Uncertainty Range Fitting for $C_\text{HII}$}
\tablehead{
 \colhead{Model} & \colhead{$C_0$} & \colhead{$\alpha_0$} & \colhead{$\alpha_1$} &\colhead{$\beta_0$}
}
\startdata
$C_{\text{HII},\Delta<50}$ & [1.7,2.4] & [-1.7,-1.3] & [-0.28,-0.17] & [-1.3,-1.3]\\
$C_{\text{HII},\Delta<100}$ & [2.1,3.2] & [-2.2,-1.9] & [-0.27,-0.17] & [-1.3,-1.4] \\
$C_{\text{HII},\Delta<200}$ & [2.4,4.4] & [-2.7,-2.5]  & [-0.22,-0.18] & [-1.3,-1.4]
\enddata
\end{deluxetable}

One key term in the global reionization equation is the  clumping factor in the recombination term. In Section \ref{sec:balance}, we have defined recombination clumping factor $C_\text{R}$ with spatial variation of recombination rate. In practice, sometimes the recombination rate $\alpha$ is treated as a constant, and its value is fixed at the averaged temperature of the medium and is independent of the recombination case. In this case, the recombination clumping factor is reduced to the ionized hydrogen clumping factor:
\begin{equation}
\label{eq:cHII}
    C_{\text{HII}} = \frac{\langle n_{\text{HII}}^2 \rangle_{\rm{V}}}{\langle n_{\text{HII} } \rangle_{\rm{V}}^2}.
\end{equation}

Previous work have also provided fits for the ionized hydrogen clumping factor and used them to solve for the evolution of the ionization fraction. For example, \cite{Shull2012a} uses hydrodynamic simulations with a uniform ionizing background to fit $C_\text{Shull}$, corresponding to our $C_{\text{HII}}$, and their fitting is used in \cite{Madau2017a} for computing the volume-filling factor $
Q$, which we will discuss in more detail in Section \ref{sec:xi}. \cite{Kaurov2015a} also makes measurements of ionized hydrogen clumping factor using Radhydro simulations and applies lower limit of $\sim 0.9$ on the ionization fraction in their calculations to only account for ionized gas. Another common practice in the literature is to use a constant $C=3$ model for the recombination term \citep[e.g.][]{Kuhlen2012a,Robertson2015a, Bouwens2015a}, mostly based on the two simulation measurements mentioned above. In this section, we show our measurements from the RadHydro simulations as an alternative to the existing models which use uniform photoionizing backgrounds.

We first measure the ionized hydrogen clumping factor $C_{\text{HII}}$ defined in Equation \ref{eq:cHII}. As said in Section \ref{sec:clumping}, we make our measurements with three overdensity thresholds $\Delta = $ (50, 100, 200) to exclude the dense regions around the halos. However, unlike many previous works \citep[e.g.][]{Jeeson-Daniel2014a,Kaurov2015a}, we do not impose a lower limit on the overdensity, nor do we have a lower limit on the ionization fraction of the cell. This is because we are using the clumping factors more for the purpose of global analytical modeling of reionization history rather than subgrid models within simulations.

Inspired by the fitting function of $C_{\text{H}}$, we use the same functional form, a single power law with running exponent, for the redshift dependence. While $C_{\text{H}}$ only depends on redshift, we include an additional dependence on $x_{\text{HII}}$ for $C_{\text{HII}}$:
\begin{equation}
\label{eq:cHII_fit}
    C_{\text{HII}}(z,x_{\text{HII}})  = \left[ 1+C_0 \left( \frac{1+z}{9} \right)^{\alpha(z)} \right] \langle x_{\text{HII}} \rangle_{\rm{M}} ^{\beta_0}
\end{equation}
where $\alpha(z) = \alpha_0+\alpha_1(z-8)$, and $C_0$, $\alpha_0$, $\alpha_1$ and $\beta_0$ are parameters 
that we will find via optimization. Here the ionized hydrogen clumping factor depends on both redshift and mass-weighted ionization fraction because we want to allow for different reionization modeling when using our fits.

Similar to the fitting method introduced in the previous section, we use the basin-hopping algorithm and minimize the relative error. Here, instead of fitting the data from one simulation, we now optimize against Sims 0, 1, and 2 simultaneously to account for the ionization fraction dependence, and we fit different sets of parameters for the three density cuts. Here we choose $z_0$ in Equation \ref{eq:error} to be the redshift when $\langle x_{\text{HII}}\rangle_{\rm M} = 1\%$ and $z_{\rm{n}}$ to be the redshift when $\langle x_{\text{HII}}\rangle_{\rm M} = 99.9\%$, which result in different redshift ranges for the three simulations. The best fit parameters are listed in Table \ref{tab:CHII}. From the table, we see that the general trend in the three parameters for the redshift dependency part is similar to that in $C_{\rm H}$. As for ionization-fraction dependency, we see that the models for three density cuts all have $C_{\rm HII} \propto x_{\rm HII}^{-1.3}$, so the dependence on ionization fraction is not affected much by the density upper limit.

The resulting evolution of $C_{\text{HII}}$ matches simulation results within $20\%$ for any models and definitions, and has mean square errors of within $5\%$. With all parameters monotonic in the density cut $\Delta$, one could interpolate these parameters for intermediary definitions of the clumping factor $C_{\text{HII}}$ for further simulations.

Figure \ref{fig:cHII} shows the ionized hydrogen clumping factors from the three simulations and the fitted curves for the three simulations. $C_{\text{HII}}$ starts off at a relatively high value of $\sim 10^3$ due to the early ionization of the high density regions around the sources and then decrease rapidly with redshift as larger regions get ionized. Towards the end of reionization, almost all of the neutral hydrogen are ionized, so the value of $C_{\text{HII}}$ becomes close to the total hydrogen clumping factor at order unity.

To demonstrate uncertainties in $C_{\text{HII}}$, we also plot in Figure \ref{fig:cHII} the estimated error bands to the clumping factors based on the differences between the mid-resolution and high-resolution simulations. From the plots, we see that there is a $\sim20\%$ difference between different resolutions. We also fit the upper- and lower-limits of the clumping factors and provide fitting parameters in Table \ref{tab:CHII_range}, in order to show how the range of parameters corresponds to the range in clumping factors.

Figure \ref{fig:cHII_rho} shows the dependence of $C_{\rm HII}$ on the density cuts. Similar to $C_{\rm H}$, lower density cuts (e.g. $\Delta<50$) exclude more regions around the halo and result in a lower clumping, which also leads to a lower total recombination rate. The lower panels show the deviations of the fits from the simulation data, and we can see that with the fitting formula given in Equation \ref{eq:cHII_fit}, we can fit the three simulations within $5\%$ error during most of reionization, although the fitting can get larger than data values at high redshift due to the limitation of the functional form.

Comparing to previous works on $C_{\text{HII}}$ \citep[e.g.][]{Pawlik2009a,Shull2012a,Jeeson-Daniel2014a,Kaurov2015a}, our measured values are higher at $z>8$, and the reason are two-folds: firstly, our simulations use radiative transfer to track the evolution of photons and gases, so the process is patchy throughout reionization. This leads to higher patchiness, and a higher $C_{\text{HII}}$,comparing with the simulations that turn on a uniform ionizing background around $z=8$. Secondly, as mentioned before, we did not apply any lower limit on either $\Delta$ or $x_{\text{HII}}$. This means that we have included all cells outside halos, leading to a larger $C_{\text{HII}}$ at high redshift when the ionization fraction is very inhomogeneous.

\subsection{Recombination Clumping Factor}
\label{sec:ceff}

\begin{figure*}[t]
\includegraphics[width=1.0\hsize]{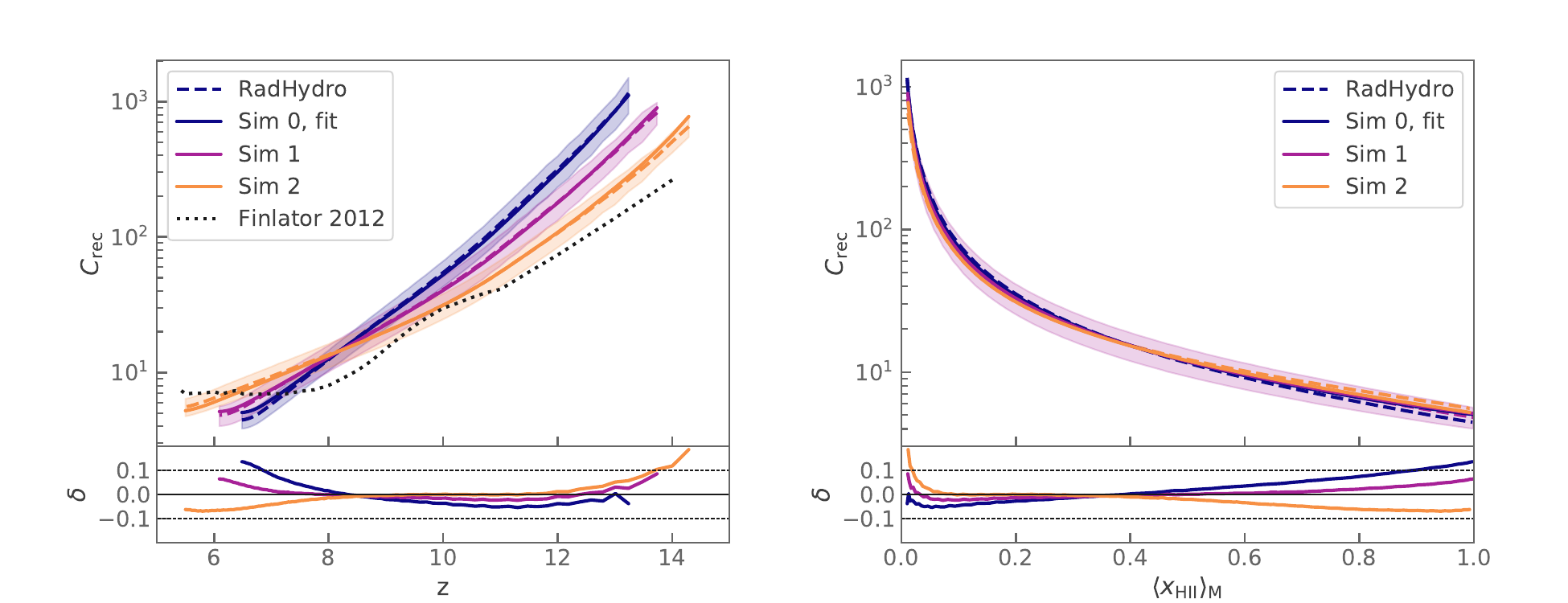}
\caption{Measurements of the recombination clumping factors from the three RadHydro simulations as a function of redshift with the $\Delta<100$ density cut. We show the simulation data (\textbf{dashed lines}) together with estimated uncertainties (shaded regions) and our fitting results (\textbf{solid lines}) for Sims 0 (\textbf{blue}), 1 (\textbf{purple}) and 2 (\textbf{orange}). We also show the results from \cite{Finlator2012a} (\textbf{black dotted}) for comparison. The lower panel shows the percentage difference between the fitted results and the simulation results. In most periods throughout reionization, our fits are within $10\%$ of the simulation results.} 
\label{fig:ceff}
\end{figure*}

\begin{figure*}[t]
\includegraphics[width=1.0\hsize]{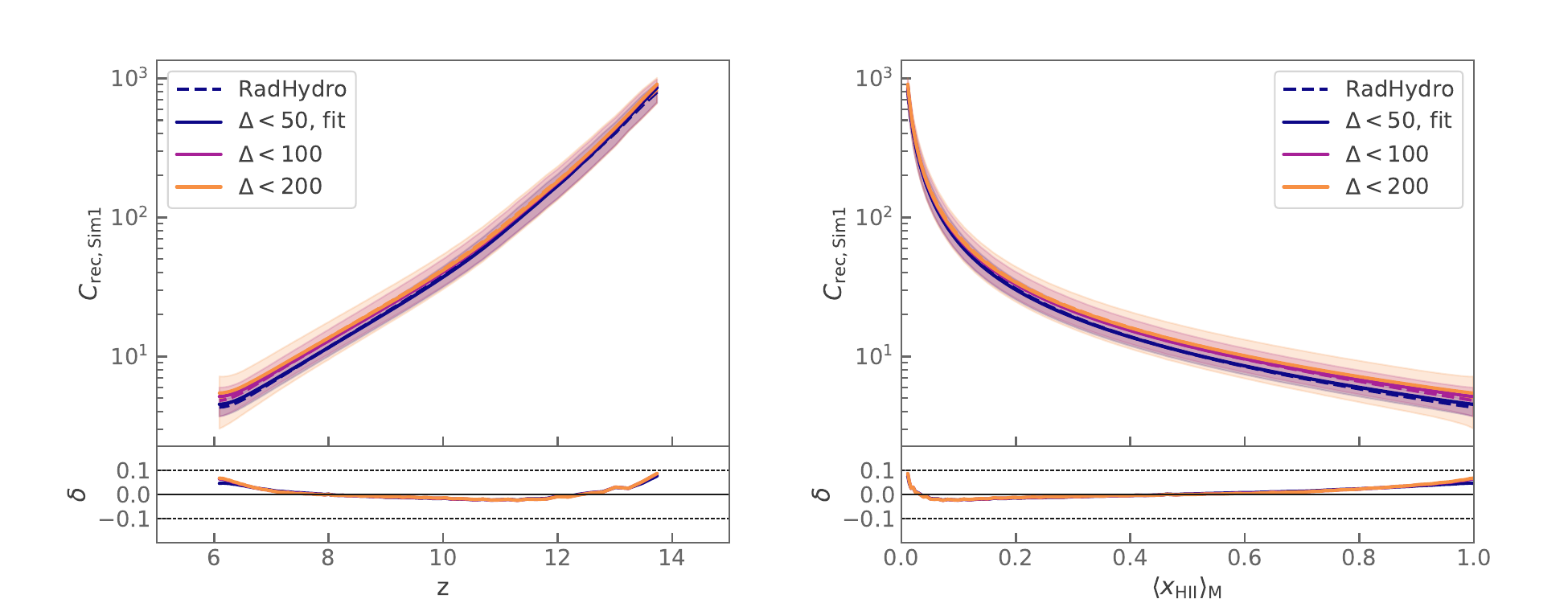}
\caption{ The effect of different density cuts on the recombination clumping factor within Sim 1. The effect of density cuts on $C_{\rm rec}$ is not very prominent compared to that on $C_{\rm HII}$.} 
\label{fig:ceff_rho}
\end{figure*}

\begin{deluxetable}{lCCCCCC}
\label{tab:CR}
\tablewidth{\textwidth}
\tablecaption{$C_\text{rec}$: Normalized Recombination Clumping Factor}
\tablehead{
 & & & $C_\text{rec}$\\
\hline
\colhead{Model} & \colhead{$C_0$} & \colhead{$\alpha_0$} & \colhead{$\alpha_1$} &\colhead{$\beta_0$} &\colhead{$\delta_\text{rms}$} & \colhead{$\delta_\text{max}$}}
\startdata
$C_{\text{rec},\Delta<50}$ & 2.96 & -1.31 & -0.310 & -1.39 & 4.9\% & 14\%\\
$C_{\text{rec},\Delta<100}$ & 3.38 & -1.50 & -0.341 & -1.41  & 2.0\% & 8.6\%\\
$C_{\text{rec},\Delta<200}$ & 3.57 & -1.58  & -0.354 & -1.42 & 3.7\% & 19\%
\enddata
\end{deluxetable}

\begin{deluxetable}{lCCCC}
\label{tab:CR_range}
\tablewidth{\textwidth}
\tablecaption{Uncertainty Range Fitting for $C_\text{rec}$}
\tablehead{
 \colhead{Model} & \colhead{$C_0$} & \colhead{$\alpha_0$} & \colhead{$\alpha_1$} &\colhead{$\beta_0$}
}
\startdata
$C_{\text{rec},\Delta<50}$ & [2.4,3.5] & [-1.5,-1.2] & [-0.31,-0.32] & [-1.4,-1.4]\\
$C_{\text{rec},\Delta<100}$ & [2.6,4.1] & [-1.7,-1.4] & [-0.32,-0.36] & [-1.4,-1.4] \\
$C_{\text{rec},\Delta<200}$ & [2.7,4.5] & [-1.7,-1.6]  & [-0.32,-0.38] & [-1.4,-1.5]
\enddata
\end{deluxetable}

\begin{figure*}[t]
\includegraphics[width=1.0\hsize]{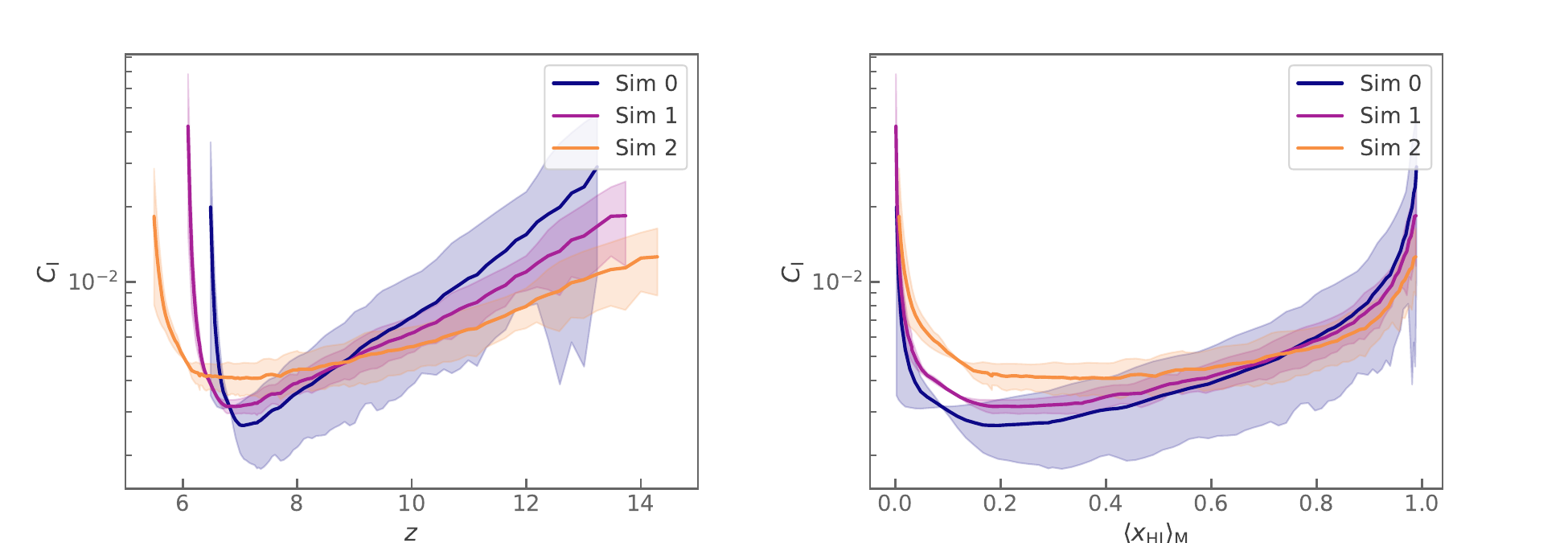}
\caption{The evolution of the photoionization clumping factor $C_{\text{I}}$ as a function of redshift (\textbf{left}) and mass-weighted neutral fraction (\textbf{right}) for three RadHydro simulations. Same as in Figure \ref{fig:cH}, the estimated uncertainties are shown by the shaded region.}
\label{fig:cHI}
\end{figure*}

As mentioned in Section \ref{sec:balance}, instead of assuming a constant recombination coefficient, it is more accurate to take the spatial variation of $\alpha$ into account when calculating the recombination clumping factor. Hence, we also calculated $C_{\text{R}}$ defined in Equation \ref{eq: Ceff} from our three simulations and we provide our results and fits here. In our calculation, one slight difference from the definition in Equation \ref{eq: Ceff} is that instead of using $\langle \alpha \rangle$ in the denominator, we choose a fiducial value of $\alpha = \alpha_{\rm{B}}(2\times 10^4K)$, such that our renormalized recombination clumping factor $C_{\rm rec}$ is defined as:
\begin{equation}
    C_{\text{rec}} \equiv \frac{\langle \alpha(T) n_{\text{e}} n_{\text{HII}} \rangle}{\alpha_{\rm{B}}(2\times10^4K) \langle n_{\text{e}} \rangle\langle n_{\text{HII}} \rangle}
\end{equation}
In practice, this is only a matter of multiplication by a constant, and we find it more convenient to choose a widely used value. The readers can easily rescale $C_\text{rec}$ differently if they want.

Figure \ref{fig:ceff} and Figure \ref{fig:ceff_rho} shows our measurements of $C_{\text{rec}}$ from the RadHydro simulations, for three simulations and three density cuts respectively. Same as $C_{\text{HII}}$, we also plot the uncertainty estimation from resolution effects with the shaded bands. The general trend and magnitude resembles the ionized hydrogen clumping factor. Therefore, we use the same functional form (Equation \ref{eq:cHII_fit}) as in $C_\text{HII}$ to fit the recombination clumping factor:
\begin{equation}
\label{eq:cR_fit}
    C_{\text{rec}}(z,x_{\text{HII}})  = \left[ 1+C_0 \left( \frac{1+z}{9} \right)^{\alpha(z)} \right] \langle x_{\text{HII}} \rangle_{\rm{M}} ^{\beta_0},
\end{equation}
where $\alpha(z) = \alpha_0+\alpha_1(z-8)$, and $C_0$, $\alpha_0$, $\alpha_1$ and $\beta_0$ are parameters we want to optimize.

Like in $C_{\rm HII}$, we also fit three sets of parameters for three different density cuts, and each set of parameters is fitted to three simulations simultaneously. The parameters from the fits are presented in Table \ref{tab:CR} with uncertainty ranges shown in Table \ref{tab:CR_range}. From the tables, we see that the dependence of $x_{\rm HII}$ on redshift is similar for $C_{\rm HII}$ and $C_{\rm rec}$ and almost the same for three density cuts. The exponent in the redshift dependence has larger linear terms, leading to larger slope at high redshifts.  

For comparison, we add the data of $C_{\text{R}}$ from \cite{Finlator2012a} to Figure \ref{fig:ceff}. The data was originally in $x_{\text{HII,V}}C_{\text{R}}$ and used $\alpha(10^4K)$ for the averaged recombination coefficient (note that their $C_{\rm R}$ is defined similarly to our $C_{\rm rec}$, although the normalization is different), but to show a direct comparison with our results, we divide it by $x_{\text{HII,V}}$ and rescale to $\alpha(2\times10^4K)$. After rescaling, we see from the plot that our data and fits agrees with the data from \cite{Finlator2012a}, although there are some discrepancies due to our different reionization histories and box sizes. Reionization is later in our simulations, and the difference of ionization fractions enters into the difference in the clumping factor, where later reionization can lead to higher clumping at the same redshift. Furthermore, the box size from \cite{Finlator2012a} is 8 times smaller than ours, which leads to differences in the source distribution and can also result in our higher clumping factors comparing with theirs. Also from \cite{Finlator2012a}, we can see that if one intends to use our models for simulation sub-grid, it is possible to use $x_{\text{HII,V}}C_{\text{R}}$ instead of applying an ionization fraction lower limit at $x_{\text{HII}} = 0.95$, as the evolution of the two values are close to each other. 
\subsection{Photoionization Clumping Factor}
\label{sec:source}

To use Equation \ref{eq:Gamma} with the ionization rate $\Gamma$, we also need to compute the ionization clumping factor $C_\text{I}$ defined in Equation \ref{eq:CI}. The measurement of $C_\text{I}$ was previously computed by \cite{Kohler2007a} using a small (4 $\text{h}^{-1}$Mpc) box. In their work, the focus was towards the very end of reionization due to the interest in Lyman-$\alpha$ lines, and the evolution of ionization clumping factor in the long redshift interval before the end was ignored in the fit. In our measurement, we want to track the evolution of $C_{\text{I}}$ throughout reionization. 

In Figure \ref{fig:cHI}, we show the evolution of $C_{\text{I}}$ as a function of redshift and mass-weighted neutral fraction, together with our uncertainty estimations. From the left panel, we can see a turnover at $z = 6\sim7$, when reionization is mostly complete. Before this turnover redshift, the value of $C_{\text{I}}$ was decreasing with time, and this is because the ionization front propagates to the larger regions in the IGM where the process of photoionization became less concentrated in space. Then, towards the end of reionization, photonionization rate was once again dominated by the remaining neutral regions in the IGM close to halos where the density is high, and so the ionization clumping factor has a steep increase. The fluctuations seen in the plots are due to the episodic star formation in a finite-size box.

\subsection{Ionization Fraction}
\label{sec:xi}

\begin{figure*}[t!]
\includegraphics[width=1.0\hsize]{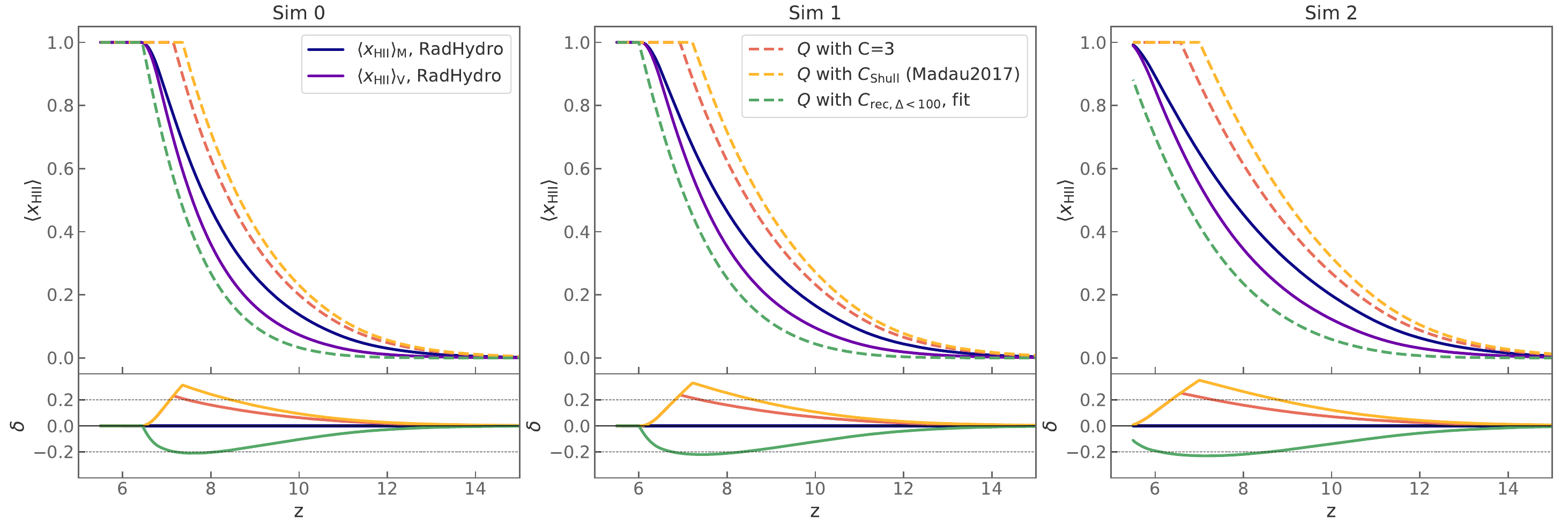}
\caption{The M99 model reionization histories in comparison with the simulation reionization histories. From left to right are the reionization histories from Sims 0, 1, and 2. We calculate the volume-filling factor Q with fixed recombination factor $C=3$ (\textbf{yellow dashed}), $C_{\rm Shull}$ from \cite{Shull2012a} following \cite{Madau2017a} (\textbf{red dashed}) and with our fitted recombination clumping factor $C_{\rm rec}$ (\textbf{green dashed}). We plot the calculated results of Q against the mass-weighted (\textbf{blue solid}) and volume-weighted (\textbf{purple solid}) reionization fractions from the RadHydro simulations. In the lower panels we show the fractional difference of the M99 histories from the simulation results.} 
\label{fig:xi_madau}
\end{figure*}

\begin{figure*}[t]
\includegraphics[width=1.0\hsize]{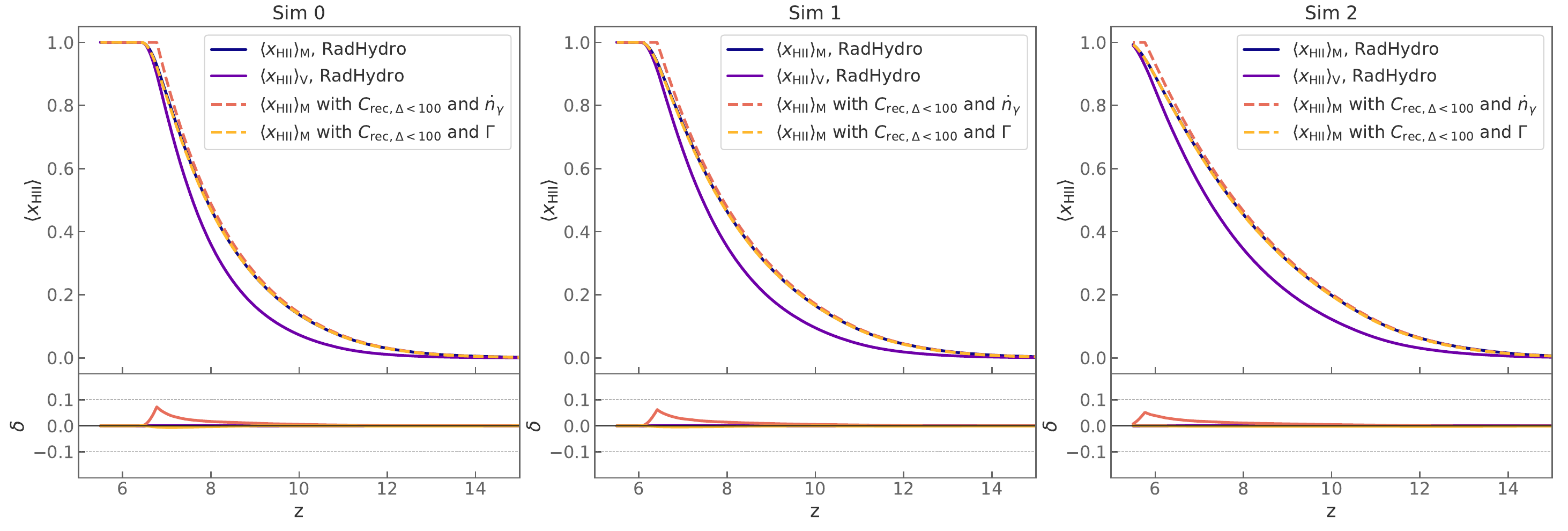}
\caption{The evolution of $\langle x_{\text{HII}}\rangle_M$ from the analytical model proposed in this work, solved with ionizing photon production rate $\dot{n}_\gamma$ (\textbf{red dashed}) and photoionization rate $\Gamma$ (\textbf{yellow dashed}). We plot them against the mass-weighted (\textbf{blue solid}) and volume-weighted (\textbf{purple solid}) reionization fractions from the RadHydro simulations. The resulting ionization fractions follow those from the simulations much closer than the M99 models.} 
\label{fig:xi_new}
\end{figure*}

Having fitted the clumping factors needed in solving the differential equation for the evolution of hydrogen ionization fraction, we now compare our models (Equation \ref{eq:Gamma} and \ref{eq:ndot}) calculated with the recombination clumping factor, to the M99 formalism calculated with constant $C=3$, the \cite{Shull2012a} fit and our fit for the recombination clumping factor $C_{\rm rec}$.

First, we calculate the evolution of volume filling factor $Q$ by solving Equation \ref{eq:m99} following \cite{Madau2017a}, where we assume an average temperature of $T=2\times10^4K$ and use $C=3$ and $C_{\text{Shull}} = 2.9 \left( {(1+z)}/{6}\right)^{-1.1}$ to model the recombination rate. We choose these two clumping factors because the $C=3$ model is a widely adopted simplification and $C_{\text{Shull}}$ reproduces the \cite{Madau2017a} calculation. For comparison purposes, we also use our fitted clumping factor $C_{\rm HII }$ together with  the M99 model to solve for the volume-filling factor $Q$, although we caution the reader that this is not a self-consistent solution. The photoionization term is approximated by $\dot{n}_\gamma$ as in M99, where $\dot{n}_\gamma$ is directly measured from the simulations (see \cite{Trac2015a} for details).

In Figure \ref{fig:xi_madau}, we show the M99 model reionization histories with $C=3$, $C_{\rm Shull}$ and $C_{\rm rec}$ in comparison with the simulation reionization histories. From left to right are the reionization histories from Sims 0, 1, and 2.  We plot them against the mass- and volume-weighted reionization fractions from the RadHydro simulations. In the lower panels we show the percentage difference of the M99 histories from the simulation results. 

To better isolate the effect of the analytical modeling from the use of different clumping factors, let us first look at the volume-filling factors solved with our fitted $C_{\rm rec}$. We can see from the plots that solving the M99 equations with $C_{\rm rec}$ results in later reionization compared to the simulations, and this is due to the fact that the M99 recombination term which is linear in $Q$ overestimates the recombination rate, and thus postpones the overall ionization of hydrogen. Now if we instead use the $C=3$ and the $C_{\rm Shull}$ clumping factors, as was done in \cite{Madau2017a}, then we would see the opposite effect of an earlier reionization compared to both our simulation results and the $z_{\rm rei}=7$ value in \cite{Shull2012a}. The main reason for this difference is that the recombination clumping factors measured from our radiative transfer simulations are much larger than those measured from a uniform ionizing background, especially during the early stage of reionization. This underestimation of recombination clumping factor over-corrects for the overestimation of recombination rates due to the linear term in $Q$. In either case, the deviation of the solved $Q$ can deviate from the simulated reionization histories by up to $20\%$.

To use the volume filling factor model to solve for the reionization history, it is necessary to redefine the clumping factor used in the recombination term. The clumping factor should be $C_{\rm R,Q} = C_{\rm R}Q$, where $C_{\rm R}$ follows the definition in Equation \ref{eq: Ceff}, in order for the recombination term to match that in Equation \ref{eq:ndot}. Based on the comparison with \cite{Finlator2012a} in Section \ref{sec:ceff}, we can see that $C_{\rm R}Q$ is close to a recombination clumping factor with a $>95\%$ ionization fraction cut, so one may also use the $C_{\rm R} (x_{\rm HII}>95\%)$ calculated within highly ionized regions together with the M99 model.

Next, we use our model in Equation \ref{eq:Gamma} and \ref{eq:ndot}, together with our fits for $C_{\rm rec}$ in Equation \ref{eq:cR_fit} to solve for the mass-weighted ionization fraction. Here we choose the $C_{\rm rec}$ with $\Delta<100$ to solve for reionization history because it is consistent with our definition of the escape fractions in the RadHydro simulations. This density cut also falls between the mean overdensity of spherical top-hat halos and the overdensity at the virial radius of an isothermal DM halo during reionization \citep{Pawlik2009a}. Using $C_{\rm rec,\Delta<50}$ instead will lead to slightly earlier reionization as the recombination rate becomes lower, and using $C_{\rm rec,\Delta<200}$ will lead to slightly later reionization. 
Note that $C_{\rm rec,\Delta<50}$ and $C_{\rm rec,\Delta<200}$ are not consistent with our reionization rates in Equations \ref{eq:Gamma} and \ref{eq:ndot}, although in either case, the effect of density cuts on solving reionization history is small, with a maximum difference of $<5\%$.

Figure \ref{fig:xi_new} shows the evolution of $\langle x_{\text{HII}}\rangle_{\rm M}$ from the analytical model proposed in this work, solved with ionizing photon production rate $\dot{n}_\gamma$ and photoionization rate $\Gamma$. The resulting ionization fractions follow those from the simulations much closer than the M99 models. In particular, the ionization fraction solved with photoionization rate $\Gamma$ shows almost no deviation from the simulation results. The model with $\dot{n_\gamma}$ also doesn't deviate much from the simulation data, with an at most $5\%$ difference in Sim 0 towards the end of reionization. 

In the above solutions we have used $\dot{n_\gamma}$ and $\Gamma$ directly from our simulations. However, for future applications people might have different $\dot{n_\gamma}$ and $\Gamma$ from ours. Our analytical model is derived from a first-principle local ionization balance equation, and so it has the generality to be applicable to different reionization scenarios. In our three simulations, for example, we also have different ionization rates and the model works fine in all three cases. One has to take caution and make sure that the photon-production rate is either defined consistently with our density cuts, or interpolated from our values shown in the tables. Also, since there is redshift and ionization-fraction dependence in the clumping factors, when using different ionization rates one has to recalculate the clumping factors at each redshift iteratively when solving for the reionization history. As long as there is consistency in the definition of all the variables in the equation, it is fine to use our global analytical equations in other reionization models.

As was mentioned in Section \ref{sec:ceff}, our clumping factors are subject to uncertainties. Therefore, we test the sensitivity of our model to different variations in the clumping factors within the uncertainties. We use the upper- and lower-limit fittings of the recombination clumping factors shown in Table \ref{tab:CR_range} to solve for the mass-weighted ionization fraction, and find that using the upper limits will result in a maximum of 4\% lower ionization fraction while using the lower-limits will result in a maximum of 4\% higher ionization fraction, when we keep the photon-production rate the same. Thus, we conclude that our analytical model is not very sensitive to the variation in clumping factors from the resolution limits.

\subsection{Thomson Optical Depth}
\label{sec:tau}

\begin{deluxetable*}{lCCCCC}
\label{tab:tau}
\tablewidth{\textwidth}
\tablecaption{Thomson Optical Depth}
\tablehead{
 & & &$\tau_{\rm{e}}$\\
\hline
\colhead{Model} & \colhead{M99, C = 3} & \colhead{M99 + Shull12} & \colhead{Our model w/ $\Gamma$} &\colhead{Our model w/ $\dot{n}_\gamma$} & \colhead{RadHydro}
}

\startdata
Sim 0 & 0.065 & 0.067 & 0.059 & 0.060 & 0.060 \\
Sim 1 & 0.066 & 0.069 & 0.059 & 0.060 & 0.060\\
Sim 2 & 0.067 & 0.070  & 0.061 & 0.059& 0.060
\enddata
\end{deluxetable*}

One goal of computing the ionization fraction of hydrogen through the analytical models is to infer the Thomson optical depth $\tau_e$, which is an important observable for reionization and parameter for cosmological surveys. It is defined as:
\begin{equation}
\label{eq:tau}
    \tau_{\rm{e}}(z) 
    = \sigma_T \int_0^{z}  dz' \frac{dt}{dz'} \langle n_{\rm e}(z') \rangle_{\rm V} ,
\end{equation}
where the volume-averaged free electron number density,
\begin{align}
    \langle n_{\rm e}\rangle_{\rm V} & = \langle n_{\rm HII}\rangle_{\rm V} +\langle n_{\rm HeII}\rangle_{\rm V} + 2\langle n_{\rm HeIII}\rangle_{\rm V} , \nonumber \\
    & = \langle x_\mathrm{HII}\rangle_\mathrm{M} \langle n_{\rm H}\rangle_{\rm V} + \langle x_\mathrm{HeII}\rangle_\mathrm{M} \langle n_{\rm He}\rangle_{\rm V} \nonumber \\ 
    &+ 2\langle x_\mathrm{HeIII}\rangle_\mathrm{M}\langle n_{\rm He}\rangle_{\rm V} ,
\end{align}
is related to the volume-averaged number densities and mass-weighted ionization fractions for hydrogen and helium. The redshift integration is performed from the present up until the beginning of reionization.

We calculate $\tau_{\rm{e}}$ using different models for the ionization fractions, and compare how choice of reionization models affects the value of $\tau_{\rm{e}}$. In our calculation, we assume single ionization of helium until redshift $z=3$, after which helium gets fully ionized. For each reionization history model, we calculate $n_{\rm{e}}$ from $\langle x_{\text{HII}}\rangle_{\rm M}$ or $Q$ (note though that the correct calculation of $n_{\rm e}$ should use the mass-weighted ionization fraction) and integrate from the beginning of reionzation down to redshift $z=0$.

Table \ref{tab:tau} shows the optical depths from four different models in comparison with the $\tau_{\rm{e}}$ values from the RadHydro simulations. Not surprisingly, just like the reionization history, the calculation based on Equation \ref{eq:Gamma} follows the simulation results most closely, but the approximated method with $\dot{n}_\gamma$ in Equation \ref{eq:ndot} also resembles the simulations in terms of $\tau_{\rm{e}}$. With the two models following M99, the values of $\tau_{\rm{e}}$ are systematically higher by $5\sim15\%$ due to the earlier reionization of the gas as already shown in Figure \ref{fig:xi_madau}. 

In addition, note that Table \ref{tab:tau} shows the total optical depth which includes both the reionization and the post-reionization contributions. Out of the two, the post-reionization term contributes $\sim 0.03$ to all of the models equally, so that if we only consider the difference between the reionization optical depth when constraining reionization models, the deviation from the simulation results could be as large as $20\% \sim 30\%$ with the M99 models.

Given the latest $\tau_{\rm{e}}$ measurement from \cite{Planck2018a} of $0.054\pm 0.007$, which has a $13\%$ uncertainty, conclusions about the IGM using the optical depth with the M99 model could be systematically biased, while the error in $\tau_{\rm{e}}$ calculated with our model is well within the tolerance of current observations.

\section{Conclusion}
In this paper, we address issues in the analytical models proposed in \cite{Madau1999a} based on Str{\"o}mgren-sphere analysis, and propose an alternative model to solve for mass-weighted ionization fraction based on the local ionization balance equations. There are two main differences between our model and the M99 model: firstly, our derivation leads to a recombination term quadratic in the ionization fraction, while the M99 model uses a linear term; secondly, we do not assume constant clumping factors of ionized hydrogen. We also do not use globally averaged temperature to calculate the recombination coefficient $\alpha(T)$, although this has a minor effect in solving for the reionization history compared with the previous differences. 

To study the evolution of clumping factors and include spatial fluctuations into the recombination term, we use the results of the RadHydro simulations from the SCORCH project. The measurements are based on three RadHydro simulations described in detail in \cite{Doussot2019a}. These simulations assume identical initial conditions and parameters except for the evolution of the ionizing photon escape fractions. The escape fractions $f_{esc}$ are constant, linear and quadratic in the three simulations, leading to different reionization histories. To match the idea of temporal and spatial variation of clumping factors and recombination coefficients in the analytical model we propose, those three simulations do not compel the clumping factor to be constant but rather record its free evolution as another output result.

Our first key result is the time-evolution of the ionized-hydrogen clumping factor $C_{\text{HII}}$. We show that $C_{\text{HII}}$ depends both on redshift and on the mass-weighted ionization fraction. We measured $C_{\text{HII}}$ from the three RadHydro simulations using three different definitions, setting the over-density cut at $\Delta<50$, $\Delta<100$ and $\Delta<200$ to exclude the regions within the halos. We clearly see the strong dependence of $C_{\text{HII}}$ on redshift, where $C_{\text{HII}}$ starts off at a very high value due to the patchiness of reionization, and drop down quickly to order unity towards the end of reionization. Based on the data from all three simulations, we provide empirical fitting functions of $C_{\text{HII}} (z,\langle x_{\text{HII}}\rangle)$, and show that it fits the measured $C_{\text{HII}}$ to within $20\%$ throughout reionization. 

Next, we take into account the spatial variation of temperature and measure the normalized recombination clumping factors $C_{\text{rec}}$ in the same way as $C_{\text{HII}}$. We also provide fits for $C_{\text{rec}}$ and the fits are within $20\%$ relative error from the simulation results. With caution, our fit can be interpolated for other empirically set density limits and, to some extent, other reionization histories.

In addition to the clumping factors in the recombination term, we also show the redshift dependence of the ionization clumping factor $C_{\text{I}}$ and total-hydrogen clumping factors. In particular, we pay attention to the entire redshift range throughout reionization, instead of focusing on the end. Our measurements of both result in higher values comparing to previous works, and we observe a turning point in the evolution of $C_{\text{I}}$ with the neutral fraction.

Finally, we use both \cite{Madau2017a} methods and our models and fits to solve for the evolution of mass-weighted ionization fraction, and compare both to our simulation results. While M99 model results in a $>20\%$ difference from all three simulations, our model fits the simulation results to within $5\%$. The low clumping factor used in \cite{Madau2017a} results in an earlier end of reionization, and it also has a $>10\%$ difference in the Thomson optical depth $\tau_{\rm{e}}$ and a $>20\%$ difference in $\tau_{\rm{e,reion}}$. Our model, on the other hand, matches both the ionization history and the value of $\tau_e$ from the simulations much better, with a $<5\%$ difference from simulations. 

\acknowledgements
We thank the anonymous reviewer for asking thoughtful questions and providing suggestions to make our work more comprehensive. We thank Alexander Kaurov, Francois Lanusse, and Michelle Ntampaka for helpful discussions. We are grateful for the data provided by Kristian Finlator in his paper \cite{Finlator2012a}. N.C.~is supported by John Peoples Jr.~Research Fellowship at Carnegie Mellon University. A.D.~acknowledges the McWilliams Center for Cosmology for hosting his internship. H.T.~acknowledges support from STScI grant HST-AR-15013.002-A.
R.C.~is supported in part by NASA grant 80NSSC18K1101. Simulations were run at the NASA Advanced Supercomputing (NAS) Center.

\bibliography{scorch3}{}
\bibliographystyle{aasjournal}

\end{document}